\documentclass[journal]{IEEEtran}
\usepackage{afterpage}
\usepackage{amssymb}
\usepackage{amsmath}
\usepackage{amsfonts}
\usepackage{array}
\usepackage{authblk}
\usepackage[english]{babel}
\usepackage{booktabs}
\usepackage{color}
\usepackage{courier}
\usepackage{diagbox}
\usepackage{enumerate}
\usepackage{epsfig}
\usepackage[T1]{fontenc}
\usepackage{graphicx}	
\usepackage{helvet}
\usepackage[utf8]{inputenc}
\usepackage{mathrsfs}
\usepackage{mathtools}
\usepackage{multirow}
\usepackage[square,sort,comma,numbers,compress]{natbib}
\usepackage{overpic}
\usepackage{setspace}
\usepackage{stmaryrd}
\usepackage{siunitx}
\usepackage{subfigure}
\usepackage{threeparttable}
\usepackage{times}
\usepackage{url}
\usepackage{varwidth}
\usepackage{xpatch}
\usepackage[usenames,dvipsnames]{xcolor}
\usepackage{float}
\usepackage{algorithm}
\usepackage{algorithmicx}
\usepackage{algpseudocode}
\usepackage{bm}
\usepackage{hyperref} 
\usepackage{amsmath}
\allowdisplaybreaks[4]
\usepackage{amssymb,amsmath}

\xpatchcmd{\citeauthor}{\begingroup}{\begingroup\em}{}{}

\xpatchcmd{\citet}{\begingroup}{\begingroup\em}{}{}
\begin{document}

\renewcommand{\figurename}{Fig.}
\markboth{IEEE Transactions on Cybernetics}
{Shell \MakeLowercase{\textit{et al.}}: Bare Demo of IEEEtran.cls for IEEE Journals}


\title{A Dynamic Systems Approach to Modelling Human-Machine Rhythm Interaction}

\author{Zhongju Yuan, Wannes Van Ransbeeck, Geraint Wiggins\textsuperscript{*}, Dick Botteldooren\textsuperscript{*}

\thanks{This work was supported by the BOF grant (BOF/24J/2021/246) and the WithMe FWO grant (3G043020). This research was also partially supported by the Flemish AI Research Programme. (\emph{Corresponding author: Geraint Wiggins and Dick Botteldooren})}
\thanks{Z. Yuan, W. Van Ransbeeck, and D. Botteldooren are with WAVES Research Group, Ghent University, Gent, Belgium (e-mails: zhongju.yuan@ugent.be; wannes.vanransbeeck@ugent.be; dick.botteldooren@ugent.be).}
\thanks{G. Wiggins is with AI Lab, Vrije Universiteit Brussel, Brussel, Belgium \& EECS, Queen Mary University of London, London, UK (e-mail: geraint.wiggins@vub.be).}
\vspace{-0pt}}
\maketitle

\newcommand\blfootnote[1]{%
\begingroup
\renewcommand\thefootnote{}\footnote{#1}%
\addtocounter{footnote}{-1}%
\endgroup
}

\def\ie{\emph{i.e.}}
\def\eg{\emph{e.g.}}
\def\cf{\emph{cf.}}
\def\etal{{\em et al.}}
\def\etc{{\em etc.}}

\newcommand{\para}[1]{\vspace{.05in}\noindent\textbf{#1}}
\newcommand{\revise}[1]{{\color{green}{#1}}}
\newcommand{\zk}[1]{{\color{blue}{[Revise:#1]}}}
\newcommand{\todo}[1]{{\color{red}{#1}}}
\newcommand{\green}[1]{{\color{black}{#1}}}
\newcommand{\red}[1]{{\color{black}{#1}}}
\newcommand{\blue}[1]{{\color{black}{#1}}}

\begin{abstract}
In exploring the simulation of human rhythmic perception and synchronization capabilities, this study introduces a computational model inspired by the physical and biological processes underlying rhythm processing. Utilizing a reservoir computing framework that simulates the function of cerebellum, the model features a dual-neuron classification and incorporates parameters to modulate information transfer, reflecting biological neural network characteristics. Our findings demonstrate the model's ability to accurately perceive and adapt to rhythmic patterns within the human perceptible range, exhibiting behavior closely aligned with human rhythm interaction. By incorporating fine-tuning mechanisms and delay-feedback, the model enables continuous learning and precise rhythm prediction. The introduction of customized settings further enhances its capacity to stimulate diverse human rhythmic behaviors, underscoring the potential of this architecture in temporal cognitive task modeling and the study of rhythm synchronization and prediction in artificial and biological systems. Therefore, our model is capable of transparently modelling cognitive theories that elucidate the dynamic processes by which the brain generates rhythm-related behavior.
\end{abstract}

\section{Introduction}\label{SECIntro}

\IEEEPARstart{T}{here} are a lot of works are working on the modelling cognitive tasks~\cite{yang2019task, jaffe2023modelling, ritz2024orthogonal}, specifically on the rhythmic perception task~\cite{vuust2014rhythmic}. Rhythm is a time series that exhibits periodic patterns or cycles~\cite{brown2004smoothing} over a certain time interval. The sounds that humans use for communication are temporally structured sequences of events, such as musical notes. Here, {\it rhythm} means the pattern of timing and stress in the amplitude envelope of an acoustic sequence~\cite{large2015neural}, which has a basic beat, the {\it tactus}, in a specific frequency range. The {\it meter} is a perceived temporal structure that includes the tactus frequency~\cite{feld1984generative, london2012hearing}. 

In humans and embodied systems, the act of striking a rhythm has to begin prior to the real beats to account for delays in the system. Hence, for synchronization with others~\cite{palmer2022we}, precise prediction is crucial~\cite{koelsch2019predictive, heggli2019musical}. Humans are capable of rhythm and rhythm synchronization within a particular range of beat and meter frequencies, a characteristic that we want the model architecture to simulate. In accordance with the process of biological rhythm processing in music, we designed a reservoir containing a pool of resonators to fulfill a role similar to the cerebellum~\cite{kudithipudi2022biological, o2022critical}. Reservoir computing is commonly used for time series prediction tasks~\cite{xu2016adaptive}. In the work by Rigotti et al.~\cite{rigotti2013importance}, the authors present a biological model of cognition where single-neuron activity in the human brain is tuned to combinations of multiple task-related aspects, resembling the nature of the reservoir readout layer.

We propose a physical inspired reservoir to model a predictive coding task~\cite{rigotti2013importance, yang2019task}. In the proposed reservoir structure, neurons are classified into two types: primary and intermediate. The primary neurons correspond to pressure in the inspirational physical system, while the intermediate neurons map to particle velocities. To modulate information transfer, we introduce two types of weights, $c$ and $k$. In the equivalent physical system, they represent propagation speed and decay rate, respectively. This physical inspiration imposes certain constraints on the reservoir weight matrix, yet it also enables easy adjustments to the overall reservoir speed and activation level. Furthermore, distinct regions within the reservoir can be designated distinct roles (e.g., oscillation frequencies), allowing for targeted design of their dynamics.

The biological plausibility of the proposed architecture could also be examined. Traveling waves and oscillations play a crucial part in dynamically coordinating neural connectivity, allowing for the flexible organization of the timing and directionality of network interactions throughout the cortex, which is essential for supporting cognition and behavior~\cite{mohan2024direction}. The overall topology-preserving structure can also be observed in the auditory cortex, where specific areas exhibit greater responsiveness to sound modulated at various frequencies~\cite{janata2002cortical,schonwiesner2009spectro}. At the level of individual biological neurons, the intermediate artificial neurons can be considered as (groups of) synapses with a temporal storage capacity. This temporary information storage capability enables the intermediate neurons to model delays, a crucial aspect of biological neural networks~\cite{madadi2018dendritic}. The response of the primary neurons is further modulated by a global parameter that may be analogous to neurotransmitter control. This combination of structures enables the generation of slow rhythms without compromising the biological constraints of individual biological neurons. Additionally, the double-parameter architecture can discriminate between low and high spontaneous activity neurons, which exhibit varying sensitivities~\cite{taberner2005response}.

For the basic task of rhythm perception, specifically the predictive coding task related to meter, this study conducted tests using a single input. Our model is designed to exhibit a range of rhythm frequency perception closely aligned with human capabilities~\cite{large2008resonating, london2012hearing}, capable of perceiving frequencies within the human perceptible reservoir. Rhythms beyond the frequency perception of humans—either too slow or too fast—are similarly challenging for the model to perceive. By adjusting global variables, we can emulate human-like behaviors, such as accelerating or decelerating in response to perceiving the beat too early or too late. Furthermore, through the dynamic selection (DS) mechanism, it is possible to adjust the activated portions within the reservoir in real-time, allowing the readout layer to focus more on areas within the reservoir that are of greater significance for the predictive coding task.

We aspire for the model to not only "perceive" meter but also to exhibit human-like behavior during interactions with another system (human or machine) in a poly-rhythmic way (e.g. 2 against 3 rhythm). As the other system may drift or intentionally change, we have incorporated a fast adaptation mechanism to facilitate continuous learning within the model~\cite{matthews2022perceived, kudithipudi2022biological}. Additionally, we introduced delayed feedback to transform the model into a closed-loop system, mirroring human capability to receive feedback while tapping to a rhythm. Such feedback also facilitates the model to become generative once the external stimulus has faded away. 

To explore how human-like the proposed model behaves in an interaction task, it was tested against a human experiment where either a human interacts with a perfect rhythm generated by a computer or with another human. The model exhibited similar imperfections as human subjects during interactions with a computer. However, due to the lack of customization within the model, it was challenging to replicate the diverse rhythms produced by different subjects under varying circumstances. To address this, we introduced customized settings for each participant in subsequent human-to-human interactions. These settings were tailored to define whether an individual was more likely to adhere to induced musical rhythms or be influenced by the other participant. 
\section{Methods}\label{Methods}
The main objective of the current model is to predict the occurrence of rhythmic beats during interactions with human-like (im)precision in a biologically plausible way. 

The proposed model uses reservoir computing to capture the temporal aspects of rhythms. Reservoirs typically exhibit multiple damped resonances. Because slow temporal behavior can be obtained based on interactions between neurons that are fast, this can be considered a biologically plausible backbone. Our innovation contains two steps. Firstly, we introduce a topology-preserving reservoir structure (Section \ref{section:structure}). During training it is combined with classical output weight training. Secondly, the new structure allows tuning the reservoir during the application phase by means of parameters with physically predictable impact that are biologically interpretable. Based on our specifically designed synchronization loss function and the Dynamical Selection (DS) mechanism, adjustments are made to the reservoir's connection matrix to fine-tune predictions. This ensures accurate matching between the predicted output and the desired target, as detailed in Section \ref{section:synchronization}. In order to enhance the model's capacity to accurately learn new tasks, we introduce a rapid adaptation mechanism, which is shown in Section \ref{section:fast_adaptation}.

\subsection{A Framework for motor-auditory interaction}

\begin{figure*}
    \centering
    \includegraphics[width=\textwidth]{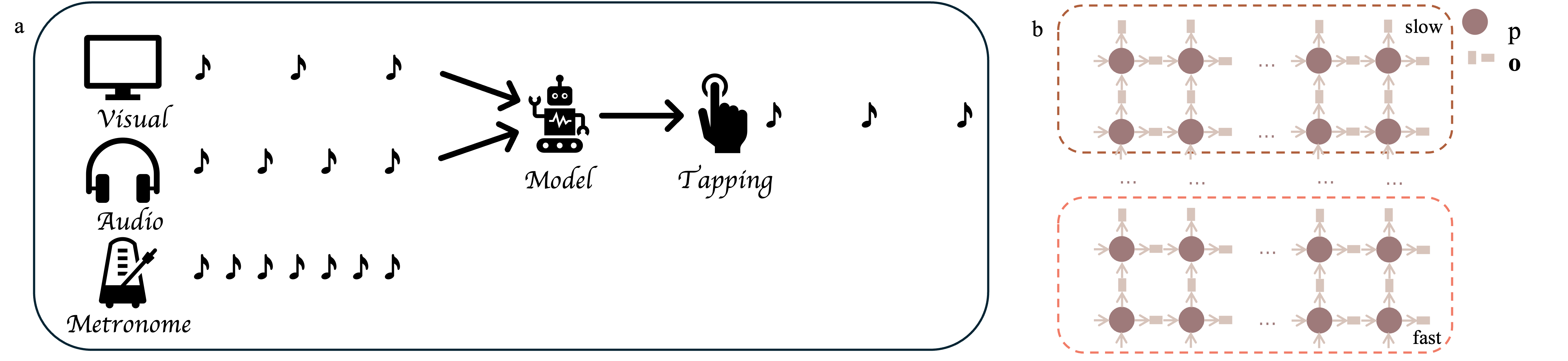} 
    \caption{\textbf{Illustration of a typical task for the model and the reservoir structure.} Panel (a) shows a typical task: the model is primed with one rhythm (noted as visual here, referring to the human priming further on in this paper) while exposed to another rhythm based on the same meter; its task is to predict the beats of the primed rhythm and continue doing so after the primer fades. Panel (b) shows the proposed reservoir structure. It illustrates the components $p$, and $\mathbf{o}$ of the reservoir.}
    \label{fig:overview}
\end{figure*}

A typical task for our model is illustrated in Fig \ref{fig:overview}: beat a rhythm based on another aurally-presented rhythm with the same underlying meter. To instruct the model what rhythm it is required to beat, a priming phase is foreseen. As this will be done via visual stimulation in the human experiment we refer to it as the visual input in Fig \ref{fig:overview}, yet as far as the model is concerned it consists of a second periodic signal envelope. Computer models are inherently very fast and could simply follow, but for the system to be human-like and even implementable in a robot it needs to be predictive. Here we selected a 200 milliseconds forward prediction. 

To create this predictive behavior, an oscillating system is needed. A reservoir of connected neurons is chosen as a complex resonating system containing thousands of degrees of freedom. To select the desired rhythm, the internal states of this system are combined using trainable output weights. Training is done during the priming phase and could represent the biological process of selecting the appropriate envelope following component from the disentangling of the rhythm by the auditory system.

The outcome of this training process is a generative model designed to simulate sequential human behavior in rhythmic cognitive tasks. Central to the generative model is a latent dynamical system characterized by state variables $\mathbf{h}_t$.  The evolution of the latent state is governed by the following dynamics:
\begin{equation}
    \mathbf{h}_{t+1} = f_{\theta}(\mathbf{h}_{t}, \mathbf{x}_{t}, \mathbf{b}_{t}, \xi_{t}),
\label{eq:reservoir}
\end{equation}
where $\mathbf{x}_{t}$ represents the inputs (primer, auditory reference, feedback) at time $t$, $\mathbf{b}_{t}$ is the bias at time $t$, $\xi_{t}$ is the noise term at time $t$, and $f_{\theta}$ denotes the dynamics function. Here, $f_{\theta}$ is modelled as a highly expressive physical-inspired reservoir structure (Section \ref{section:structure} ).

The representation of the input employs a smooth pulse to accentuate the occurrence or lack of a beat at any given moment. The injection of zero-mean Gaussian noise $\xi_{t}$ at each time step is essential. From a physical perspective, the noise will trigger the oscillating system to exhibit "natural" metronomes. Additive noise is biologically plausible due to spontaneous emission of the (auditory) neurons.

Model outputs, denoted by $\mathbf{y}_{t}$, are produced through a linear combination acting as a representation of mixed selectivity~\cite{rigotti2013importance}, derived from the activity of neurons within the reservoir. The motor behavior is not explicitly modelled yet it is assumed that this introduces a delay of $\Delta t$. Therefore, the model is trained to predict upcoming beats  $\Delta t$ ahead of their occurrence. For training the output layer weights, we employ the Mean Squared Error (MSE) metric to ascertain the proximity of the model's predictions to the target behavior.

\subsection{Reservoir weights}
\label{section:structure}

The proposed model is based on a reservoir computing (RC) method: the Echo State Network (ESN)~\cite{jaeger2007echo, 10491369}, a specific recurrent neural network~\cite{verstraeten2007experimental, jaeger2004harnessing,maass2002real}. Its hidden states change according to the current input and the hidden states from the previous time step, which follows the equations: 
\begin{equation}
\begin{aligned}
    \mathbf{h}_{t+1} &= (1-\alpha)\mathbf{h}_{t}\\
    &+\alpha f(\mathbf{W}_{in}\mathbf{x}_{t}+\mathbf{W}\mathbf{h}_{t}+\xi_{t}), \\
    \hat{\mathbf{y}_{t}} &= \mathbf{W}_{out}\mathbf{h}_{t},
\end{aligned}
\label{eq:model}
\end{equation}
where $\mathbf{W}$ is a sparse matrix defining the connectivity of the network, $\mathbf{W}_{in}$ is the input weight, and $\mathbf{W}_{out}$ is the output weight matrix, and $\alpha$ is the leakage rate of the model. $f(\cdot)$ is a non-linear function, for which $tanh(\cdot)$ is used in this paper. $\mathbf{x}_{t}$ is the input signal at time step $t$, and $\mathbf{h}_{t}$ is the hidden state at time step $t$, and $\hat{\mathbf{y}_{t}}$ is the output of the model at time step $t$, which is the prediction. The expected prediction should satisfy $\mathbf{y}_t=\mathbf{x}_{t+n}$ where n is the number of time steps predicted ahead: $\Delta t = n \delta t$ with $\delta t$ the simulation time step. 

Conventionally, the weight matrix of the reservoir $\mathbf{W}$, the input weight matrix $\mathbf{W}_{in}$, and the bias matrix $\mathbf{b}_{t}$ are randomly generated at each time step. In this paper, $\mathbf{W}_{bias}$ includes the Gaussion noise $\xi_{t}$, regenerated every time step. The weight matrix $\mathbf{W}$ is usually adapted to keep all of its eigenvalues inside the unit circle~\cite{manneschi2021exploiting} in the z-domain, thereby assuring stability of this dynamic system in the linear, low amplitude, regime.

For the problem at hand, predicting the occurrence of rhythmic beats in a human-like way, the poles in the z-domain of the $\mathbf{W}$ matrix and thus the resonances in the random reservoir are not optimally placed: (1) they span a frequency range that does not match human capabilities; (2) many of them are too much damped. To overcome this problem, we propose a novel reservoir structure designed following physical principles. To simplify the tuning of $\mathbf{W}$, we design it based on a 2D Finite-Difference Time-Domain (2D-FDTD) computational approximation of the linearized Euler equations~\cite{botteldooren1995finite} for wave propagation in a medium with randomly generated properties. Because this system results in local connections, it has a clear topology which allows crafting connections from input or outputs to areas showing specific dynamics. Starting from the wave equations Eq ~\ref{eq:wave_eq} where $c$ is the wave speed and $k$ is a damping (amplification if negative) factor, and $p$ and $\mathbf{o}$ are proportional to pressure and velocity. 
\begin{equation}
\begin{aligned}
    \frac{\partial p}{\partial t} + c^2 \nabla \mathbf{o} &= 0 \\
    \frac{\partial \mathbf{o}}{\partial t} - k \mathbf{o} + \nabla p &= 0,
\end{aligned}
\label{eq:wave_eq}
\end{equation}

The simplest FDTD model, a staggered grid, central differences, and an explicit time stepping approximation of these equations leads to their discretised form (Eq ~\ref{eq:wave_eq_disc}):
\begin{equation}
\begin{aligned}
    p_{i,j}(t+\delta t) &= p_{i,j}(t) + c_{i,j}^{2}\delta t * (o_{x,i+1,j}-o_{x,i,j})/\delta x \\&+ c_{i,j}^{2} \delta t * (o_{y,i,j+1}-o_{y,i,j})/\delta y
     \\
    o_{x,i,j}(t+\delta t/2) &= {1-k_{i,j}\delta t/2\over{1+k_{i,j}\delta t/2}} o_{x,i,j}(t-\delta t/2) \\&+ {\delta t \over{ \delta x (1+k_{i,j}\delta t/2)}} (p_{i,j}-p_{i-1,j}) 
    ,
\end{aligned}
\label{eq:wave_eq_disc}
\end{equation}
where the indices, $i$ and $j$ refer to spatial locations and the time dependence has been omitted on the right hand side of the equation. A similar equation holds for $o_{y}$. 
Stability is guaranteed by keeping the Courant number, which relates the $\delta t$ to $\delta x$ and $\delta y$, smaller than 1.

The two groups of unknowns could be interpreted as two types of artificial neurons in a reservoir as in Fig \ref{fig:overview}: one is the primary neuron denoted as $p_{i,j}$, and the other is the intermediate neuron, labeled as $o_{x,i,j}$ or $o_{y,i,j}$ . These can be grouped in a hidden state matrix $\mathbf{x}$ like in Eq  \ref{eq:model}. 
As $p$, $\mathbf{o}$ are coupled locally and sparsely the coupling matrix $\mathbf{A}$ derived from Eq  \ref{eq:wave_eq_disc} will also be sparse. 

The weight matrix $\mathbf{W}$ of the reservoir is computed by:
\begin{equation}
\mathbf{W} = (\mathbf{A} - (1 - \alpha) \cdot \mathbf{I}) / \alpha,
\end{equation}
where $\mathbf{I}$ is the identity matrix. In this way the update equations of the reservoir Eq ~\ref{eq:reservoir} become very similar to the FDTD update equations. It implies very strong symmetry constraints on the $\mathbf{W}$ matrix. The local value of $c$ determines how strongly the $p$-neuron responds to inputs from surrounding $o$-neurons and together with the coupling to its neighbors this can result in local resonances, where the physics equivalent learns that small $c$ correspond to low-frequency resonances. By introducing a gradient in c on top of the random value, a slow (low-frequency resonances) and fast (high-frequency resonances) end of the reservoir can thus be realised as Fig \ref{fig:overview}. The variable $k$ determines how much information is transferred between the $p$ neurons that the $o$-neuron connects. Increasing $k$ will result in more strongly damped resonances.

\subsection{Tuning for synchronization}
\label{section:synchronization}

In this study, we employ stochastic gradient descent (SGD) to minimize the Mean Squared Error (MSE) between the prediction $\hat{\mathbf{y}}$ and target $\mathbf{y}$ signals during training that is based on a large number of rhythmic beats that could theoretically be encountered in music. Following training, the output layer can thus identify a suitable combination of correct oscillators, thereby providing an initial estimate of the target beat periodicity and timing.

To synchronize the prediction with the target beats more accurately, an adaptation phase is introduced. A prediction of an upcoming beat can fail in to ways: 'too early' or 'too late', hence the error is split in two parts. In both cases, there is usually an overlap between the sound envelopes corresponding to a beat in the prediction and target. If the data consists of discrete moments in time, the peak is artificially extended. Thus, the slope of the peaks is employed to calculate the error $I_{early}$ and $I_{late}$ of the prediction $\hat{\mathbf{y}}$ and target $\mathbf{y}$ signals. 

If the prediction is descending while the target is ascending, we consider the prediction to be too early. Otherwise, if the prediction is ascending while the target is descending, the prediction is too late. Both values, $I_{early}$ and $I_{late}$ until an  $\text{update\_step}$ is reached and the reservoir weights are changed, and reinitialize to 0 when the interval ends, as shown in Algorithm~\ref{alg:update_c}. 
To ensure proximity in amplitude between the target and prediction within the same time window, a moving average and a $softmax$ normalisation are first applied to both the target and prediction values:
\begin{align}
\mathbf{y}_{norm}(t) &= \frac{\mathbf{y}_{t} - \mathbf{y}_{mean}}{\mathbf{y}_{softmax}(t)}, \label{eq:norm_T}\\
\hat{\mathbf{y}}_{norm}(t) &= \frac{\hat{\mathbf{y}}_{t} - \hat{\mathbf{y}}_{mean}}{\hat{\mathbf{y}}_{softmax}(t)},
\label{eq:norm_P}
\end{align}
where
\begin{align}
\mathbf{y}_{softmax}(t) &= \ln (\int_{0}^{t} e^{\mathbf{y}_{t'}} e^{\frac{t'-t}{\tau}}dt'), \\
\hat{\mathbf{y}}_{softmax}(t) &= \ln (\int_{0}^{t} e^{\hat{\mathbf{y}}_{t'}} e^{\frac{t'-t}{\tau}}dt').
\end{align}
where $\tau$ is an exponential averaging time constant spanning multiple interbeat intervals.

By weighted comparison of $I_{early}$ and $I_{late}$, the decision is made to increase or decrease the speed-up factor $\delta_c$ by a fixed amount as shown in Algorithm~\ref{alg:update_c}. If the prediction is too early, we thus decrease all elements of $c$ proportionally to their value; if it is too late, we increase $c$ it. In this way, the whole reservoir slows down or speeds up.

Secondly, a proposed method: Dynamical Selection (DS) mechanism, is used to control the damping of the oscillations in the reservoir. The poles of the $W$ matrix are modified by identifying regions within the reservoir that are pivotal for accurately predicting beats and amplifying them by lowering $k$ in these regions. Simultaneously, the oscillations that make minor contributions are damped. To this end, each neuron within the reservoir is masked, generating masked outputs, and computed their MSE in comparison to the target in every time window. The neurons that resulted in the most significant MSE reduction when masked are considered the ones contributing the most to accurate prediction. Conversely, those neurons leading to the least reduction were considered to have the smallest contribution. We modulated the activity of these neurons by adjusting the parameter $k$ around their positions, either enhancing or diminishing their activity accordingly. As $k$ changes only slowly it will introduce some focus even after the meter or rhythmic pattern change, this mechanism can be considered to focus attention on an area in the reservoir and thus also on a specific rhythmic behavior.

\begin{algorithm}[]
\caption{Calculate loss function to adapt $c$}
\label{alg:update_c}
\begin{algorithmic}[1]
\State Init: batch\_size, $\text{update\_step}$, \text{threshold\_sum}, \text{threshold}
\For{$i$ in batch\_size}
\State $\epsilon_{early}$ = 0, $\epsilon_{late}$ = 0, $\delta_{sum}$ = 0, $\delta_{c} = 0.02$
\For{$t$ \text{in} $\text{update\_step}$}
    \If{$\mathbf{y}_{norm}(t) > \max(\hat{\mathbf{y}}_{norm}(t), 0)$}
        \If{$\mathbf{y}_{norm}(t) - \mathbf{y}_{norm}(t-1) > 0$} \If{$\hat{\mathbf{y}}_{norm}(t) - \hat{\mathbf{y}}_{norm}(t-1) < 0$}
            \State $I_\text{early}(t) = I_\text{early}(t-1) + 1$
        \EndIf
        \ElsIf{$\mathbf{y}_{norm}(t) - \mathbf{y}_{norm}(t-1) < 0$}
        \If{$\hat{\mathbf{y}}_{norm}(t) - \hat{\mathbf{y}}_{norm}(t-1) > 0$}
        \State $I_\text{late}(t) = I_\text{late}(t-1) + 1$
        \EndIf
        \EndIf
    \EndIf
    \State  $\epsilon_{early} = \epsilon_{early} +  \delta_{early} \cdot I_{\text{early}}(t)$
    \State  $\epsilon_{late} = \epsilon_{late} + \delta_{late} \cdot I_{\text{late}}(t)$
    \If{$\delta_{sum} < \text{threshold\_sum}$}
    \If{$\epsilon_{early} - \epsilon_{late} < \text{threshold}$}
        \State $\delta_{sum} = \delta_{sum} + \delta_{c}$
        \State $c *= 1 +\delta_{c}$
    \Else
        \State $\delta_{sum} = \delta_{sum} - \delta_{c}$
        \State $c *= 1 - \delta_{c}$
    \EndIf
    \Else
        \State No update.
    \EndIf
\EndFor
\EndFor
\end{algorithmic}
\end{algorithm}

\subsection{Fine-tuning and Continuation}
\label{section:fast_adaptation}
Having completed the model training, a first approximation to the output $W_{out}$ is obtained. However, predictions for task illustrated in Fig \ref{fig:overview} involving two inputs, lack accuracy. In this study, distinct combinations of beat multiples under the same meter are regarded as separate tasks. During the fine tuning process, fast adaptation is applied to the model's output layer, enabling the model to learn new combinations of multiples at different meters swiftly.

Due to the well-established initialization, the model, prior to fast adaptation, can accurately output beats at appropriate positions for tasks within the perceptible frequency range for humans. However, the amplitudes of these beats are inaccurately predicted, exhibiting significant errors compared to the target beats. Additionally, predictions sometimes include extra beats between two adjacent target beats, further contributing to discrepancies between predictions and targets. 

To facilitate fast adaptation of the output weights to new tasks, we store the fixed time step predictions and targets~\cite{finn2017model}. We compute the Mean Squared Error (MSE) between them and perform a few steps of weight updates in the direction that minimizes the MSE. To account for the impact of the additional peak between adjacent target beats on the computed MSE, a penalty is applied. Since the beats between input signals consist entirely of zeros, the MSE penalizes all extra beats equally. Therefore, we replace the zero segments between two adjacent beats with the negative portion of a sine wave with a 0.5 phase, matching the amplitude of the input peak.

During the fine-tuning phase, the adjustment formula for the model's output weights is:

\begin{equation}
    \mathbf{W}_{out} = \mathbf{W}_{out} - \mathrm{lr} * \frac{\partial \Delta \mathrm{MSE}}{\partial \mathbf{W}_{out}},
\end{equation}
where the $\mathrm{lr}$ is the defined learning rate.

Once the visual reference halts (Figure \ref{fig:overview}), fine-tuning ends. Subsequently, the model's prediction persists for an additional beat. Its detection initiates a feedback loop, and the prediction replaces the visual reference. Notably, there is no need to modify the input weight, as shown in Eq \ref{eq:feedback}. Consequently, the closed-loop model acquires the capacity to learn from its own sound and sustain the acquired rhythm.
\begin{equation}
\begin{aligned}
    \mathbf{h}_{t+1} &= (1-\alpha)\mathbf{h}_{t}\\
    &+\alpha f(\mathbf{W}_{in}(\mathbf{x}_{t}+\mathbf{\hat{y}}_{t-n})+\mathbf{W}\mathbf{h}_{t} + \mathbf{b}_{t}+\xi_{t}), \\
    \mathbf{\hat{y}}_{t} &= \mathbf{W}_{out}\mathbf{h}_{t},
\end{aligned}
\label{eq:feedback}
\end{equation}
where $\mathbf{y}(t-n)$ denotes the feedback from the output, and $\Delta t = n \delta t$ indicates that the model is predicting $\Delta t$ ms into the future.

\subsection{Customization during human behavior simulation}
When two humans interact musically, they can adopt different behaviors. In broad terms they can act as a follower or as a leader in the interaction. When comparing the proposed model with interaction experiments (Fig  \ref{fig:interaction_procedure}(b)), this is included by customizing the update learning rate. To this end, we utilize the Wasserstein distance $W(\cdot, \cdot)$ between the inter-beat interval generated by Participant 1 (P1) and the primer interval, and between the inter-beat interval generated by P1 and participant 2 (P2).

The Wasserstein distance \cite{vallender1974calculation} measures the amount of 'work' required to move one set of timings to another set, effectively identifying the inter-beat interval distribution similarity. The equation is as follows:
\begin{equation}
        W(P_{pred}, P_{target}) = \min_{\gamma \in \Gamma(P_{pred}, P_{target})} \int c(d_{pred}, d_{target}) \, d\gamma,
\end{equation}
where $P_{pred}$ and $P_{target}$ are the inter-beat interval distribution of human prediction and visual reference respectively, $d_{target}$ represents the set of inter-beat intervals from the participant, $d_{pred}$ denotes the set of inter-beat intervals from the reference or another subject, \( \min_{\gamma \in \Gamma(P_{pred}, P_{target})} \) signifies the minimum over all joint distributions \( \gamma \) with marginals \( P_{pred} \), \( P_{target} \), \( c(d_{pred}, d_{target}) \) denotes the time difference between point \( d_{pred} \) and point \( d_{target} \), and \( d\gamma(d_{pred}, d_{target}) \) represents the joint distribution.

The update learning rates are defined as:
\begin{equation}
    \beta * \frac{1}{W(P_{pred}, P_{target})}* \exp(-\frac{ds}{l}),
\label{eq:w_distance}
\end{equation}
where $\beta=0.001$ is the learning rate, $ds=6ms$ is the downsample factor, and $l$ is the sequence length.

\section{Results}\label{Results}

\subsection{The proposed model shows human-like rhythm analysis}

In human perceptive beat, lower-level dynamics, such as the grouping of two or three beats, are processed by sensory tissues such as the cerebellum, which establishes the fundamental rhythm in interaction with the environment. Meanwhile, higher cognitive processes are responsible for selecting and refining complex tasks~\cite{kudithipudi2022biological}. This allows humans to generate rhythms within a certain frequency range, and upon mastery, anticipate beats ahead of the actual timing. Slower rhythms will be subdivided into faster meters.

\begin{figure*}
    \centering
    \includegraphics[width=\textwidth]{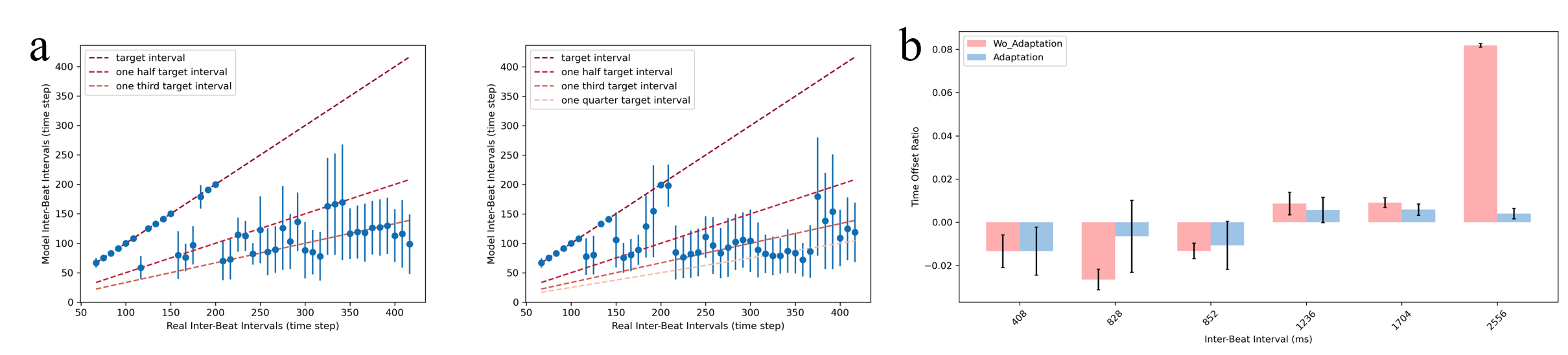} 
    \caption{\textbf{The proposed reservoir demonstrates internal meter perception.} (\textbf{a}) Comparison of the difference in inter-beat interval mean and standard deviation before and after adaptation of global parameters. (\textbf{b}) Comparison of the time offset ratio before and after adaptation of global parameters.}
    \label{fig:single_channel_results}
\end{figure*}

To assess the model's frequency perception range, the model is pre-trained on large dataset of artificially created plausible rhythms. Then, test are conducted using rhythms spanning inter-beat intervals from 400 to 3000 milliseconds with an intervals of 50 milliseconds. The time-step $\delta t$ was set to 6 milliseconds. The model was tested without fine-tuning and with and without tuning c and k for synchronisation. 

As depicted in Fig  \ref{fig:single_channel_results}, when the inter-beat interval is less than 2000 milliseconds, the model demonstrates accurate rhythm prediction, with minimal variance observed between predicted beats. However, as shown in Fig \ref{fig:single_channel_results}(a) left side, when the intervals increase, the model's predictions begin to incorporate subdivisions, notably aligning well with intervals of one half and one third of the target intervals. Such behavior would also be expected in humans as this inter-beat interval falls well beyond human capabilities for rhythmic prediction.

Following the tuning of $c$ and $k$ for synchronisation (Fig \ref{fig:single_channel_results}(a) right side) the model becomes more precise but also generates more subdivisions. A common observation is that humans also tend to perform internal counts between consecutive beats.

At six specific inter-beat intervals, the time offset ratio between the model's outputs before and after adjustments is investigated in more detail. The examples of comparing the prediction and the target rhythmic signals' mean and variance of time offset ratio are shown in Fig \ref{fig:single_channel_results}(b). The figure illustrates that in the absence of adaptation, the rhythm's offset remains small, with a relative proportion to the inter-beat interval not exceeding 4\% except for the very long 2556 msec inter-beat interval . This observation indicates the reservoir's ability to accurately predict periodic signals. By employing the proposed synchronization algorithm to adjust parameters $c$ and $k$, we alleviate instances where the original model's output leads or lags compared to the rhythm in the target. This outcome suggests that our synchronization algorithm reliably and consistently improves the prediction of upcoming beats. Note that there is a trend from anticipation to lagging as the inter-beat interval increases. 

In the human brain, traveling waves and oscillations play a fundamental role in dynamically coordinating neural connectivity, facilitating the flexible organization of the timing and directionality of neuron interactions to support cognition and behavior \cite{mohan2024direction}. When they perform a tapping task, humans synchronize with the reference and focus on the required rhythmic patterns \cite{engel2001dynamic}. We introduce two global parameters in our reservoir system to regulate timing and directionality, respectively.

\subsection{Simulation of human-AI interaction}



\begin{figure*}[!t]
\centering
    \includegraphics[width=\textwidth]{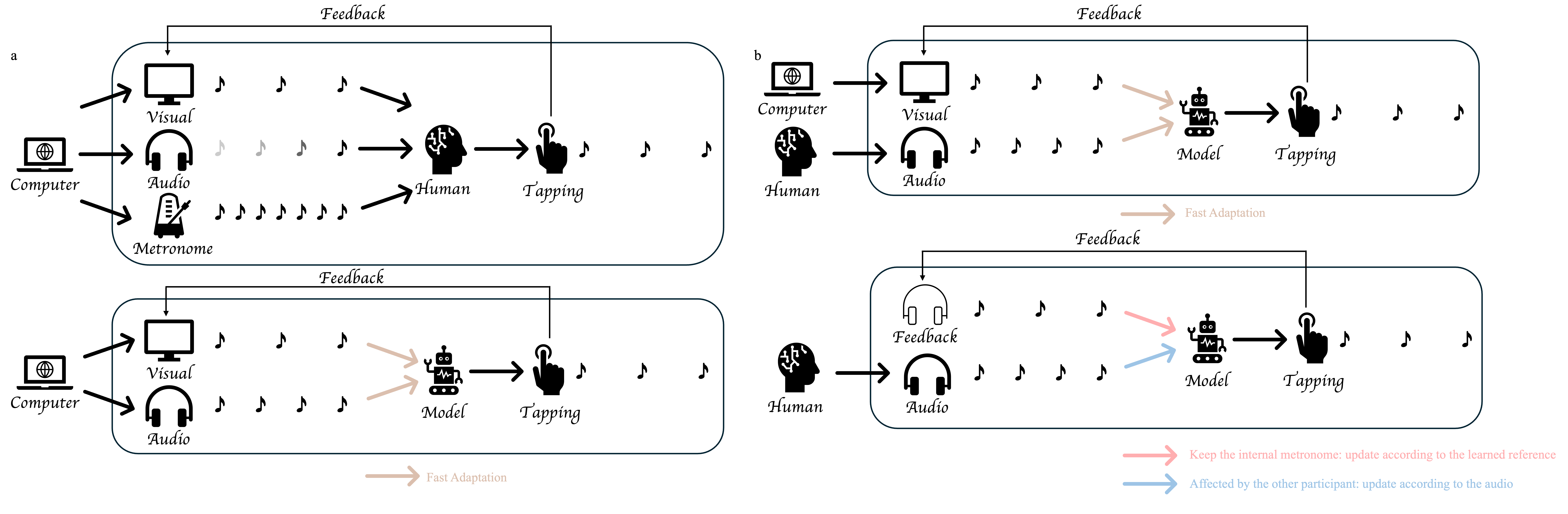}
\caption{Display of interaction procedure. (a) \textbf{Human-computer interaction experiment.} The upper part depicts the procedure of the experiment with people. During a learning phase, the participant is guided by a visual reference that gradually fades, while a different audio rhythm gradually fades in from a computer, along with a shared metronome. The participant is expected to tap the rhythm in sync with the visual reference, even after the latter has stopped. The lower part outlines the procedure for applying our model to a similar experimental setting. Given the model's proven metronome perception, the metronome is not included as an input. During learning, the input mirrors the human learning procedure, while the output weight undergoes \textbf{fine-tuning} to the dual input. After the visual reference stops, the model \textbf{feeds back} its own tapping. (b) \textbf{Human interaction procedure.} The upper part illustrates the model output weight updating procedure, which remains consistent with the human-computer interaction. The lower part demonstrates the customization added to the model's output weight. The model is affected by itself and the other channel to varying degrees.}
\label{fig:interaction_procedure}
\end{figure*}

\begin{figure*}
    \centering
    \includegraphics[width=\textwidth]{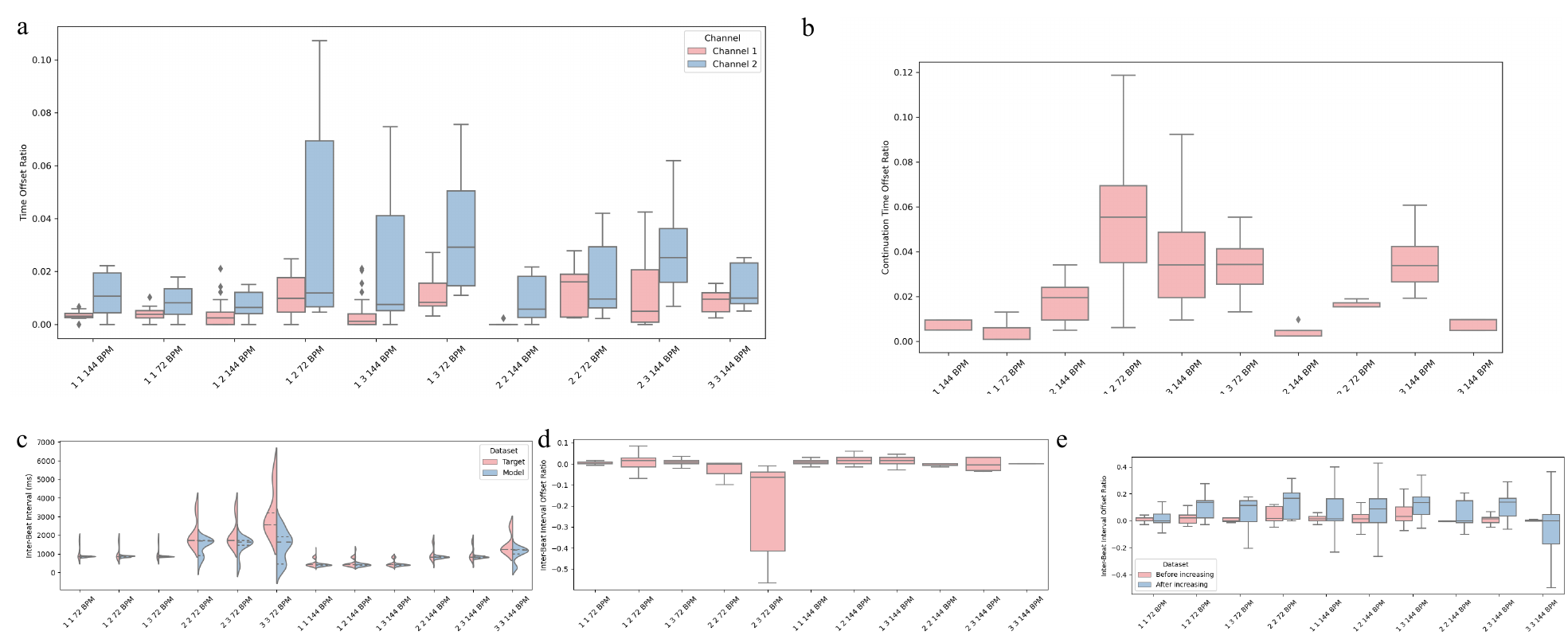} 
    \caption{\textbf{The proposed reservoir can interact with a computer similarly to a human.} (\textbf{a}) Time offset ratios for both channels in all combinations are compared. (\textbf{b}) The effect of closed-loop on the continuation task for channel 1 in all samples. (\textbf{c}) Comparison of the distribution between the skip-one target inter-beat interval and the prediction's distribution for all samples. (\textbf{d}) Illustration of the skip-a-while prediction's inter-beat interval error ratio for all samples. (\textbf{e}) After channel 2 increases the inter-beat interval by 2\%, a comparison of the prediction's inter-beat interval error ratio before and after the increase for all samples.}
    \label{fig:daul_channel_results}
\end{figure*}

As depicted in Fig \ref{fig:interaction_procedure}(a), during the initial 30 seconds of the experiment, the model receives visual input from a screen (channel 1), audio input from an audio (channel 2), and the metronome as input stimuli. All of these references originate from a computer, with the duration between beats being precisely fixed. It's important to note that the visual reference persists for 30 seconds, while the metronome gradually fades away after approximately 15 seconds. Additionally, the audio of the second subject begins to fade in initially, although they will fully perceive it after a couple of seconds. 

To further examine whether our model can effectively interact with another rhythm, we introduce dual-channel input based on the same metronome but with different beat grouping, denoted as $i:j$, where $i, j = 1, 2, 3$. We select two metronomes: one with a low frequency of 72 BPM and another with a high frequency of 144 BPM. We consistently configure the first channel to cease earlier than the second channel, designating the second channel as the other signal from another participant or computer and the first channel as the reference with a specific meter ratio.

In our model, we input the visual reference and audio from a computer as two channels with pulses in a signal, omitting the basic metronome. The target for prediction is to move the input $\Delta t$ forward. Subsequently, this target is utilized for several seconds to update the output weight, allowing the model to learn the combination of rhythms.

We classify various time interval ratios derived from the same metronome as distinct tasks. During the training process, we obtain a favorable initialization for the output weights. This initialization enables the model to generate peaks for prediction in simpler tasks, such as the 1:1 ratio, which involves a frequency easily discernible by humans. However, the predictions lack sufficient accuracy due to challenges in achieving the target amplitude. Furthermore, in samples with complex ratios, accurately predicting both channels proves difficult, resulting in the output peaks following the target instead of making predictions.

To expedite the model's learning of new tasks, we implement Fast Adaptation on the output weights based on the Mean Squared Error (MSE) between the prediction and the target. The adaptation of the output weights occurs over a certain number of time steps, with updates limited to a few steps and a small learning rate. This restriction is imposed to control the update time and computational resources within a relatively smaller range.

The time offset errors for both channels of all samples are directly compared before and after Fast Adaptation in Fig \ref{fig:daul_channel_results}(a). The figure illustrates the mean errors and variance of the errors for all peaks in each channel. Samples involving 3 multiples at 72 BPM, which are challenging for both humans and our model, exclude the 2:3 and 3:3 under 72 BPM cases. It is evident that following Fast Adaptation of output weights, the time offset errors are significantly reduced, and the predictions become more stable, as evidenced by the reduced variance across all samples. Additionally, the mean errors remain below $\Delta t$, indicating precise predictions by the model.

To enhance the model's performance to be more human-like, we introduce another task: the continuation task. This task involves participants learning to sustain a rhythmic pattern once the reference in the first channel halts. In our experimental setup, after the cessation of the reference in the first channel, the prediction accurately continues for one additional beat. Subsequently, this prediction is fed back to the model with a $\Delta t$ delayed input. In Fig \ref{fig:daul_channel_results}(b), the continuation inter-beat interval errors for all samples are compared, with mean errors consistently below 58ms. Relative to the target inter-beat interval, most errors hover around 2\%. For more intricate samples, such as 1:2 or 2:3, errors tend to be around 5\%. Thus, the closed-loop mechanism empowers the model to replicate the acquired rhythm.

The approach proposed can be conceptualized as a challenge within the realm of human-computer interaction. To theoretically gauge our model's capacity in potential real-life applications, we describe three distinct scenarios: such as the human reference skipping a beat; the human excluding portions of a reference; and the computerized metronome enhancing the inter-beat interval by 2\% with every beat. Note that, these changes start after the fast adaptation.

To evaluate whether the model has learned to predict, we employ the skip-one and skip-a-while tasks. If the model can accurately fill the skipped beats or time intervals in a predictive manner, it indicates that our model has indeed learned the rhythm. Additionally, during human and model interaction, human behavior typically exhibits some variability. It is challenging to expect all beats generated by humans to have precise timing. Therefore, introducing some variability in channel 2 by increasing the inter-beat interval by 2\% with every beat and observing the resulting behavior change in channel 1 would be reasonable.

It is important to note that due to the low frequency of the metronome, beats occurring every 3 times the metronome interval may be challenging to perceive as rhythmic. Consequently, in samples such as 3:3 72 BPM, accurately predicting both channels becomes difficult. Moreover, samples like 2:3 72 BPM perform even worse due to the inherent difficulty of the 2:3 combination task. However, 1:3 72 BPM performs relatively better, primarily because channel 1 can act as a metronome, aiding channel 2 in counting beats.

Following Fast Adaptation, we input the first channel with a signal that skips one beat every two beats, while keeping the second channel unchanged. As shown in Fig \ref{fig:daul_channel_results}(c), the inter-beat interval distribution exhibits two peaks, representing the basic interval and double basic interval. However, the distribution of the model's predicted inter-beat interval only displays the basic interval. This indicates that the model has not learned the skip-one pattern, resulting in it naturally filling the gap.

When skipping input for the first channel temporarily but retaining its delayed feedback, the generated time intervals during the pause do not deviate significantly. With the exception of outliers in each sample and the most challenging sample involving the second channel with three multiples, the majority of inter-beat interval errors are under 3\%.

After learning the provided rhythms in both channels, the inter-beat interval of the second channel begins to increase continuously and evenly. In Fig \ref{fig:daul_channel_results}(d), we compare the inter-beat interval error ratio of the first channel before and after the change in the second channel. It is evident that our model is affected by the changing channel. Prior to the increase, the prediction's inter-beat interval error is below 4\%. However, after the interval increase, the variances also escalate, particularly for high-frequency rhythms. As illustrated in Fig \ref{fig:daul_channel_results}(e), the inter-beat interval error ratio shifts towards the negative range, indicating that with the increasing inter-beat interval in the second channel, the prediction in the first channel also tends to elongate the interval. This phenomenon is observed in human behavior, where individuals tend to subconsciously adjust their pace to match that of the person they are interacting with if the other person becomes slower.

The above results are all based on scenarios that might occur in human settings. To validate the similarity between our model's outputs and human experimental results, we first conducted experiments on human-computer interaction. In these experiments, channel 2 represents the rhythm generated by the computer, which human participants listen to. Channel 1 contains a standard rhythmic reference for the first 30 seconds, followed by only their own feedback and the audio transmitted from channel 2. 

Due to the simplicity of predicting the 1:1 combination, we conducted experiments using the 2:2, 2:3, 3:2, and 3:3 combinations instead. We replaced the participants in channel 1 with our model. Subsequently, we analyzed the distribution of inter-beat intervals between the beats generated by real participants and those generated by the model. To ensure the robustness of the findings, we tested four participants.

\begin{figure*}[htbp]
    \centering
    \includegraphics[width=0.8\textwidth]{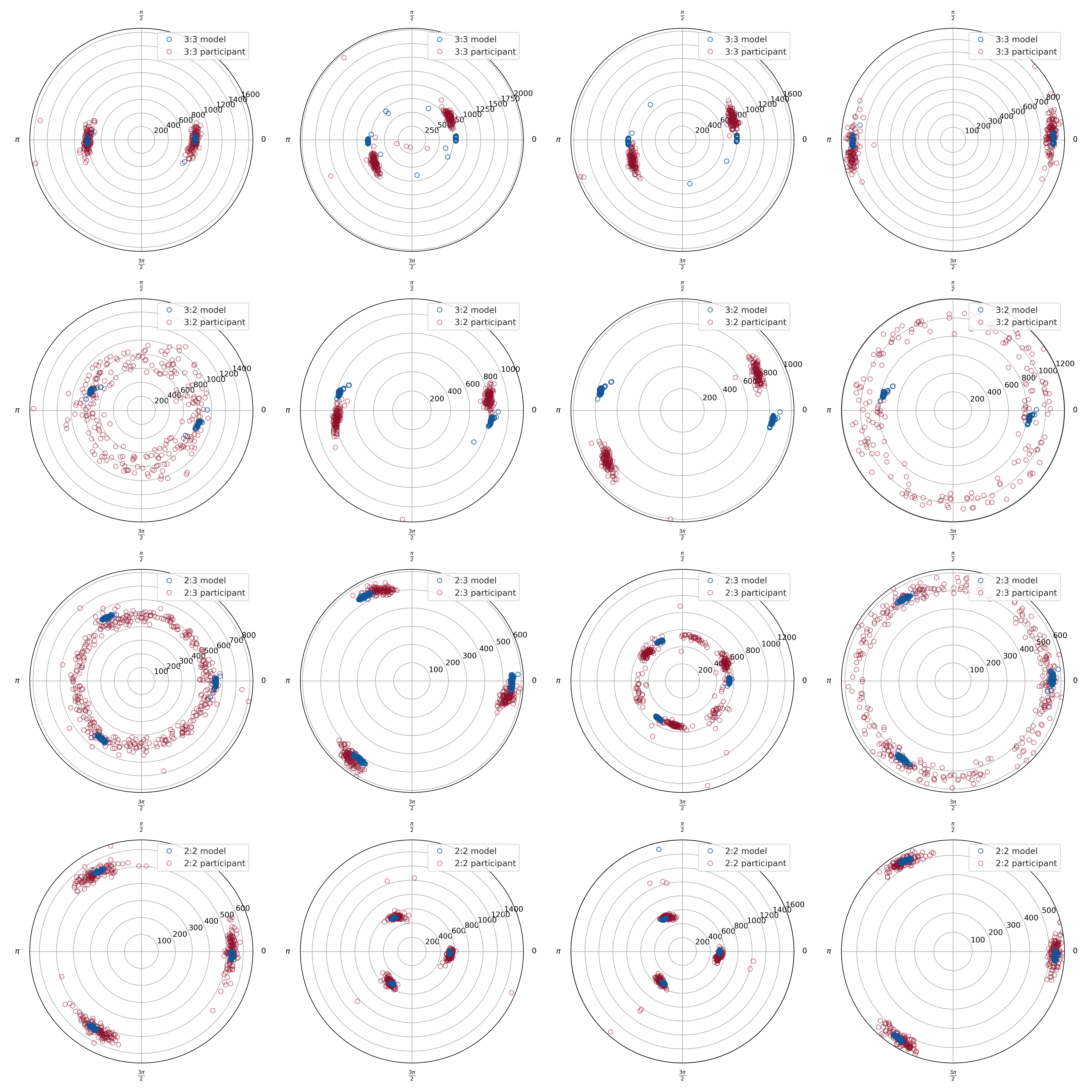}
    \caption{\textbf{Variation in Human Interaction with Identical Rhythms.} Each subplot in a row represents a different participant interacting with the same rhythm combination. The subplots in the same row correspond to the same group of participants, illustrating their varied interactions with the given rhythm. Within each subplot, four groups of participants are shown alongside their respective models. Hollow circles indicate the beats dropped by both participants and models. The angle of each hollow circle denotes the phase shift relative to the cycle (6 * metronome inter-beat interval), while the distance from the center indicates the interval between consecutive beats.}    \label{fig:huamn_computer_matching_dots_results}
\end{figure*}
\begin{figure*}[htbp]
    \centering
    \includegraphics[width=0.8\textwidth]{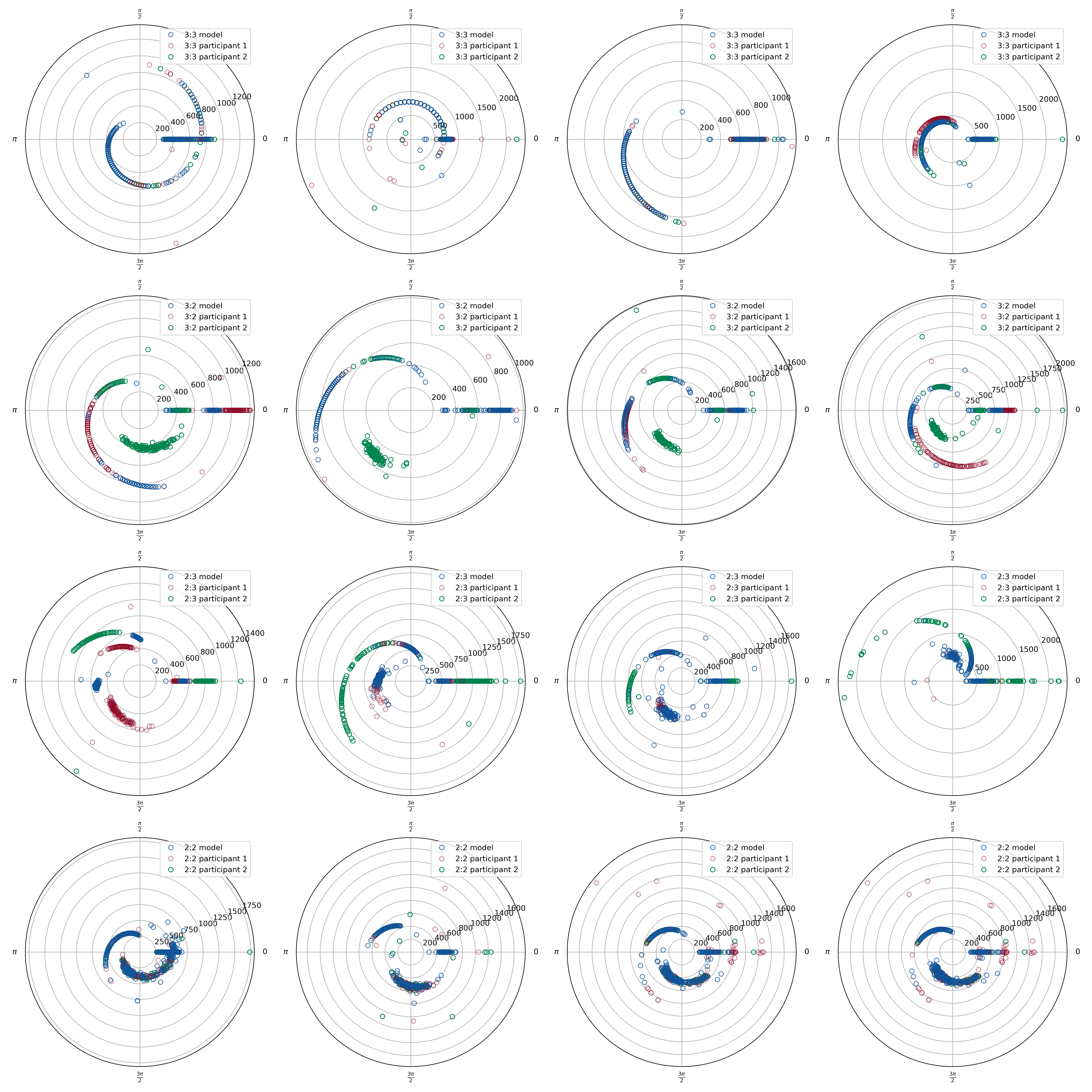}
        \caption{\textbf{Variation in Inter-Human Interaction based on different rhythm combinations.} Each subplot in a row represents a different participant interacting with the same rhythm combination. The subplots in the same row correspond to the same group of participants, illustrating their varied interactions with the given rhythm. Within each subplot, four groups of participants are shown alongside their respective models. Hollow circles indicate the beats dropped by both participants and models. The angle of each hollow circle denotes the phase shift relative to the cycle (6 * metronome inter-beat interval), while the distance from the center indicates the interval between consecutive beats.}   
    \label{fig:human_human_matching_dots_results}
\end{figure*}
The experimental results are illustrated in supplementary materials Fig S1. From the graph, it is evident that different participants exhibit varying performances when subjected to fixed computer audio inputs. In simpler tasks such as 2:2 and 3:3, the differences between participants are not significant, and the distributions can be maintained around the target interval. However, not every beat falls precisely on the target position, demonstrating some variance. This behavior mirrors that of our model, where the distribution's mean aligns with the target interval, albeit with less variance than that of the participants. In more complex tasks like 2:3 and 3:2, there is a greater disparity in performance among different participants. Some participants can sustain the learned rhythm even after the reference stops, while others are more influenced by the computer audio, causing the distribution peaks to shift towards the interval of the other channel. Since our model maintains the same input throughout the initial 30 seconds, the distribution does not vary significantly during this period. However, due to the inherent randomness integrated into the model, the distributions of the four instances are not entirely identical.

To visualize the behavior patterns of humans and the AI model, we plotted the event timings from each group in each scenario in a circular arrangement, as shown in Fig \ref{fig:human_human_matching_dots_results}. Hollow circles represent beats missed by both participants and models. The angle of each hollow circle indicates the phase shift relative to the cycle (6 * metronome inter-beat interval), while the distance from the center represents the interval between consecutive beats. The model demonstrates patterns with high accuracy and low variance. In the simpler cases (2:2 and 3:3), participants also capture the pattern but with greater variance. Compared to the participants, the model exhibits smaller variance and is less prone to phase shifts. In more complex cases, different participants display varying interactions under the same conditions. For example, in the 3:2 scenario, one participant tends to generate additional beats between two consecutive beats in the visual reference, while another misses some reference beats. The other two participants can capture the pattern but exhibit larger phase shifts, indicating they follow the visual references. In the 2:3 scenario, participants show better pattern recognition but tend to strike one event later and the next one earlier. As the figures show no clear tendency for human behavior patterns to drift toward audio channel patterns, it is unnecessary to adapt the model's output weight according to the other channel.

To analyze frequency interactions across various conditions between human participants and our model, we used Fast Fourier Transform (FFT) on the time series outputs. These comparisons are detailed in supplementary materials Fig S2. Participants show diverse frequency spectra in rhythmic tasks. Simpler tasks (2:2, 3:3) yield distinct frequency peaks, while challenging tasks (2:3, 3:2) result in more scattered spectra. The model's performance remains consistent across tasks. In the 2:2 and 3:3 tasks, human FFT spectra overlap significantly with the model's. However, in the 2:3 and 3:2 tasks, participants' responses vary, leading to scattered spectra. Some participants (e.g., participants 1 and 4 in 3:2) show less discernible frequency patterns, while high performers (e.g., participants 2 and 3 in 3:2) create subdivisions to maintain learned frequencies. Similar but less pronounced patterns are seen in the 2:3 task, indicating that higher frequency rhythms are easier to sustain. The model generally follows the visual reference frequency, occasionally showing subdivisions similar to those in humans.

\subsection{Human performance}

To simulate real human interaction, we replaced the computer providing the audio signal with actual human participants, as the procedure in Fig \ref{fig:interaction_procedure}(b) upper part. From the experiments involving human-computer interaction, it was evident that our model cannot simulate the variability among different participants. This is because the method of updating output weights is the same for all participants, and all participants stop after 30 seconds.

To enable our model to simulate genuine human activities, we need to establish customized learning rates for updating output weights (Fig \ref{fig:interaction_procedure}(b) lower part). The underlying principle is that participants who exhibit stronger control over the rhythm are better at maintaining their internal rhythm. Therefore, in subsequent updates, we should adjust more towards the reference in channel 1. Conversely, if a participant is more influenced by another participant, we should update more towards the input in channel 2. This approach ensures that the model adapts its learning rates based on the individual participant's ability to maintain their internal rhythm and their susceptibility to external influences from another participant.

To assess the participant's control over the rhythm, we compared their actual experimental data, encompassing all tapping inter-beat intervals, with two sets of data: the inter-beat intervals synchronized with channel 1's visual reference and those aligned with channel 2, originating from another participant's real inter-beat intervals. Through this comparison, we derived two distances, reflecting how closely the participant's taps align with each channel. These distances serve as measures of the participant's ability to control the rhythm and their susceptibility to external influences from another participant.

First, the adaptation of the output weight is divided into two parts: one is towards the channel 1 reference and the other is towards the channel 2 audio. The difference lies in the learning rate. The learning rate is decayed over time. We aim to measure the timing differences for all tapping timings in the sequence with channel 1's reference and channel 2 audio sequence. As there may be some phase problems during the measurement, we propose to employ the Wasserstein distance as the measurement. This distance quantifies the minimum amount of work needed to transform one distribution into the other, where "work" refers to the amount of distribution mass that must be moved, with the cost of moving unit mass from one point to another determined by a given distance function. Therefore, the learning rate will be larger if the distance is smaller by determining the learning rate as Eq \ref{eq:w_distance}.

The results of comparing the four participants and their corresponding performances are depicted in supplementary materials Fig S3. The model is utilized to simulate the behavior of Participant 1 (P1). The combination of two channels remains consistent with the human-computer interaction setup. There are four groups of participants, and our model simulates the behavior of P1 in each group. Across most combinations and groups, our model successfully emulates the participants' behavior. The inter-beat interval distributions between our simulation and the beats generated by the participants are similar. 

To compare the interaction patterns between individuals with different rhythms, a dot plot is shown in Fig \ref{fig:human_human_matching_dots_results}. Participants typically exhibit unpredictable and variable behaviors during each experiment. As a result, the hollow dots plotted in Fig \ref{fig:huamn_computer_matching_dots_results} do not display clear patterns, with angles distributed throughout the entire cycle. Therefore, we replaced the fixed-interval reference with Participant 2's behavior, resulting in Fig \ref{fig:human_human_matching_dots_results}, where the cycle is adapted dynamically. 

The basic metronome inter-beat interval is calculated as the average of two intervals generated by P2. For both the model and P1's behavior, we identify the closest beat generate by P2 and used its previous two intervals to calculate the cycle for both the model and P1. The cycle is updated for each event generated by the model and participants. To examine pattern shifts with the adaptive cycle, event timings are grouped according to the task, and each group is initialized as a new phase starting from 0. Consequently, clearer patterns relative to P2 are revealed in Fig \ref{fig:human_human_matching_dots_results}. Similar to the model’s behavior shown in Fig \ref{fig:huamn_computer_matching_dots_results}, when the task is 2, the dots appear around three angles: $0, \frac{1}{3}\pi$, and $\frac{2}{3}\pi$. For task 3, the dots are located around $0$ and $\frac{1}{2}\pi$. By introducing grouped phase initialization, the beats at the beginning of the cycle are consistently mapped to 0 degrees, though the intervals between them vary. As the radius of the dots at 0 degrees increases, the dots at other angles diverge away from the center, otherwise, the dots converge towards the center.

In all rhythm combinations, some participants tend to play faster than the reference, while others excel at maintaining their reference rhythm with closer mean inter-beat intervals and smaller variances. Different participants are influenced by each other in varying ways. For example, in the 3:3 combination, nearly all participants are significantly affected by each other, as evidenced by their interval distributions closely resembling one another. However, in the 2:3 group 2, P1 is not heavily influenced by the other participant, while the other participant struggles to maintain the rhythm behavior. P1 remains consistent with the learned rhythm despite external influences. Hence, employing the proposed varying degrees of updating the output weight, the model does not adhere rigidly to performance, akin to human-computer interaction. Instead, it exhibits differences across various combinations and participant groups. While the model may not replicate the exact inter-beat interval distribution of P1, it still demonstrates significant overlap. In all instances, the model's outputs exhibit mean values closely resembling P1's performance. However, during the updating of the output weight, occasional beats occurring between two beats are observed, mirroring findings from human-computer interaction experiments. This phenomenon suggests that humans may sometimes need to count metronome beats to maintain their internally learned rhythm.

In most cases, P1 exhibits three types of behavior: playing based on their own visual reference or following their partner. In the 2:2 and 3:3 cases, both the model and P1 typically display similar behaviors, as evidenced by the statistically similar inter-beat interval distributions shown in Fig \ref{fig:human_human_matching_dots_results}. In Groups 1, 3, and 4 for task 2:2, P1 tends to play faster than the visual reference after it disappears, with their partners showing similar behavior. The dots show the tendency to move towards the center. In Group 2, they are more likely to adhere to the presented visual reference, as the dots’ radium is more consistent. But the phase shift is different from each other. Consequently, there is a greater overlap between the model and both participants. 
For Task 3:3, the models in all groups exhibit similar behavior to P1, as indicated by the substantial overlap between their dots. 
In Group 1, the P1 is likely to change their behavior patterns during the experiment. In Group 2, the participant adheres more closely to the visual reference rhythm but with greater variance. In Groups 3 and 4, participants are likely to play faster during the experiment. 
In harder tasks, the interaction will be more complicated so these tasks will be analyzed separately. In the first group, P2 gradually increased their speed, while P1 exhibited fluctuations between fast and slow speeds. Consequently, the model is expected to be relatively accurate, resulting in increased phase variations. In the second group, P2 continued to increase their speed, while P1’s performance closely matched the model, although not deterministically. Both participants’ positions remained around their respective targets. In the third group, both P2 and P1 increased their speeds slightly, but neither deviated far from their targets, with the model consistently centered around the target. In the fourth group, P2’s speed increased gradually but not significantly, while P1’s speed decreased. The model, influenced by P2, also increased slightly in speed but not as markedly as P1’s decrease. In the first group, P2 exhibited fluctuating speeds, sometimes fast and sometimes slow, resulting in a radius variation around 800ms. P1 gradually increased their speed, but P2 maintained a 2:3 ratio. The model adhered to the rhythm of the visual reference, leading to a phase pattern with a certain angular deviation. Consequently, the model’s distribution around 0, $\frac{1}{3}\pi$, and $\frac{2}{3}\pi$ was more concentrated compared to P1. In the second and third groups, although the phase patterns of the model and P1 were more similar, P2’s behavior was entirely different. This difference is evident from the radius distribution at angle 0. In the second group, P2’s speed gradually decreased, deviating significantly from their visual reference. In the third group, P2’s speed increased, but with a smaller deviation from the visual reference. In the fourth group, P2 exhibited completely random behavior patterns, resulting in many outliers at angle 0. However, P1 and the model displayed a certain degree of similarity, with a higher overlap in their behavior.

The Fast Fourier Transform (FFT) is applied to outputs from both participants and our model, with comparisons shown in Fig S4 of the supplementary materials. Compared to interactions with a computer, spectra indicate increased noise levels, especially in 2:2 and 3:3 scenarios. Participants' frequency spectra vary with rhythmic tasks, revealing diverse interaction patterns. Our model exhibits broader variations due to tailored output weight updates. In familiar scenarios like 2:2 and 3:3, participants' spectra converge with the model's, though some participants adopt novel patterns, causing spectral differences. In tasks 2:3 and 3:2, participants' frequencies have clearer peaks, resulting in less noisy model spectra. The model's output mirrors its counterpart's frequencies, such as a minor peak aligning with P2 in the 2:3 task involving groups 2 and 3.

\section{Conclusion}\label{Discussion}

In this study, we developed a biologically inspired dynamic system to simulate human rhythmic cognition by integrating predictive coding, aligning it with human behavior. Unlike traditional data-intensive machine learning models, our approach employs a recurrent neural network (RNN) trained on basic rhythmic patterns, demonstrating the capability to learn new rhythmic combinations and interact similarly to humans.

Our model, designed with a weight matrix inspired by wave equation discretization, can emulate human-like rhythmic behaviors. To improve learning capacity, we introduced fine tuning on the output weights, enhancing the selection of neuron activity and enabling rapid adaptation to new rhythmic combinations. Our model maintains learned rhythms by selectively outputting and delaying feedback as input.

Interaction tests with and without customization revealed that our model generalizes well to most basic human rhythmic behaviors. Broadly, our framework can extend to other time-varying signals, offering a potential method for generating models that simulate the brain's dynamic functions and enable the emergence of rhythmic behavior.

\IEEEpeerreviewmaketitle
\bibliographystyle{IEEEtran}
\small{\bibliography{ref}}

\vspace{11pt}

\vfill

\ifCLASSOPTIONcaptionsoff
  \newpage
\fi
\end{document}